\documentclass[reprint, amsmath, amssymb,aps, prb]{revtex4-1}

\usepackage{graphicx}	
\usepackage{dcolumn}   	
\usepackage{bm}		
\usepackage{epstopdf}
\usepackage{float}

\usepackage{verbatim}
\usepackage{color}

\usepackage{array}
\newcolumntype{C}[1]{>{\centering\let\newline\\\arraybackslash\hspace{0pt}}m{#1}}

\begin{document}


\title{Polariton-Enhanced Exciton Transport}

\author{D. M. Myers}
\author{S. Mukherjee}
\author{J. Beaumariage}
\affiliation{Department of Physics and Astronomy, University of Pittsburgh, Pittsburgh, PA 15260, USA}
\author{M. Steger}
\affiliation{National Renewable Energy Lab, Golden, CO 80401, USA}
\author{L. N. Pfeiffer}
\author{K. West}
\affiliation{Department of Electrical Engineering, Princeton University, Princeton, NJ 08544, USA}
\author{D. W. Snoke}
\affiliation{Department of Physics and Astronomy, University of Pittsburgh, Pittsburgh, PA 15260, USA}

\date{\today}

\begin{abstract}

The transport distance of excitons in exciton-polariton systems has previously been assumed to be very small ($\lesssim 1~\mu$m). The sharp spatial profiles observed when generating polaritons by non-resonant optical excitation show that this assumption is generally true. In this paper, however, we show that the transport distances of excitons in two-dimensional planar cavity structures with even a slightly polaritonic character are much longer than expected ($\approx 20~\mu$m). Although this population of slightly polaritonic excitons is normally small compared to the total population of excitons, they can substantially outnumber the population of the polaritons at lower energies, leading to important implications for the tailoring of potential landscapes and the measurement of interactions between polaritons.
\end{abstract}

\maketitle

\section{Introduction}

The transport of excitons has great importance for solar energy and other optoelectronic applications. One of the limiting design constraints in organic solar cell design is the distance excitons can flow before they recombine \cite{Jha2009}. Exciton transport in organic and inorganic semiconductors is normally assumed to be very short, of the order of a micron or less. The properties of excitons at low momenta can be greatly altered, however, by coupling exciton states with the photon states in an optical resonator, to make exciton-polaritons. Exciting observations in polariton systems include Bose-Einstein condensation \cite{Kasprzak2006, Balili2007}, room-temperature lasing \cite{Christopoulos2007, Kena-Cohen2010room}, superfluidity \cite{Amo2009, Lerario2017}, and long-range ballistic flow \cite{Steger2015}; for a general review of exciton-polariton properties, see, e.g., Ref.~\onlinecite{Deng2010}.

In a typical system, strong coupling leads to two new eigenstates known as the upper polariton (UP) and lower polariton (LP). Near zone center, the lower polaritons have a very light mass ($\sim10^{-4} m_\mathrm{e}$) and short lifetime (less than a picosecond in most organics, to up to  200~ps in some inorganic structures \cite{Steger2015}). At higher momentum, this LP branch transforms continuously into a bare exciton branch. It is typical in the exciton-polariton literature (e.g. Refs. \onlinecite{Nelsen2013, Askitopoulos2015, Cristofolini2013, OstrovskayaPRA2012, KhanPRA2016}) to make a sharp distinction between polaritons near zone center and an exciton ``reservoir'' at higher momentum. These reservoir excitons are assumed to have a much heavier effective mass (of the order of a free electron) and long lifetime (of the order of nanoseconds), and are then assumed to not move significantly on the time scales of the polariton motion. Recent theoretical work using a simple model of a 1D chain of quantum emitters has already shown that excitons could move much longer distances when strongly coupled to a photon mode \cite{FeistPRL2015}. In this work, we show experimental evidence that the assumption of stationary excitons is not entirely valid, as we report evidence of long-range transport of highly excitonic lower polaritons. We must, indeed, distinguish between three populations on the lower polariton branch: light-mass lower polaritons at zone center, ``bare'' excitons at high momentum, with low speed due to their much heavier mass, and a third category of ``bottleneck excitons'', which have many of the properties of the bare excitons, but can travel much longer distances.

\section{Lower polariton properties}

The LP detuning ($\delta = E_\mathrm{cav}-E_\mathrm{exc}$) affects how ``excitonic'' or ``photonic'' the LP is; when detuning is negative, the LP is more photonic, and when the detuning is positive, it is more excitonic. At resonance ($\delta = 0$), the LP is exactly half photon and half exciton. We write the exciton fraction as $f_\mathrm{exc}$ and the cavity photon fraction as $f_\mathrm{cav}$, related by  $f_\mathrm{cav} + f_\mathrm{exc} = 1$  (see Ref. \onlinecite{Deng2010} for more details). The LP detuning is not only a function of the ground-state energies of the photon and the exciton modes; in a planar cavity it is also dependent on the in-plane momentum ($k_\|$) of the LP. For the LP mode, higher $k_\|$ always corresponds to larger exciton fraction. This means that at high enough $k_\|$, the LP mode is essentially no different from the original exciton mode. Under the conditions of non-resonant optical pumping with large excess energy, a large population (reservoir) of highly excitonic lower polaritons will be created at the location of the pump spot, some of which will cool into the low-$k_\|$, ``true'' LP states. Since this excitonic LP reservoir is at such high $k_\|$, its photoluminescence (PL) is typically unresolved by imaging optics. This is because high $k_\|$ corresponds to high angle of emission, and such collection is limited by the numerical aperture (NA) of the optics. Resonant optical pumping can be used to avoid producing this high $k_\|$ reservoir, instead producing a population at a specific $k_\|$ point in the LP mode \cite{VladimirovaPRB2010, TakemuraPRB2014}.

The mixing of the photon and exciton states, with the resulting change from very light mass to heavy, excitonic mass in the lower polariton branch, leads to an inflection point in the LP branch above which the effective mass becomes negative. The mass only becomes positive again above another inflection point at much higher $k_\|$. Since the lower inflection point usually corresponds to angles of emission of about $20^\circ$, corresponding to a modest NA of 0.34, and occupation is typically low near the inflection due to low density of states, it is a convenient point for distinguishing between two populations. In this work, we observe the polariton dispersion from $k_\| = 0$ to about $k_\| = 2k_\mathrm{inflection}$. We will consider all the emission below $k_\mathrm{inflection}$ to be ``normal'' lower polaritons, and all the emission above the inflection point to be ``bottleneck excitons''. The third category, of ``bare excitons,'' is not observable since their momenta lie even higher, outside the light cone for photon emission; their presence must be deduced indirectly.

The repulsive interactions between excitons and polaritons have proven extremely useful for forming various potential profiles \cite{Wertz2010, SanvittoNatPhot2011, Tosi2012, Cristofolini2013, Askitopoulos2015, SchmutzlerPRB2015, Liu2015, SunPRL2017}. From these studies, it can be seen that most of the excitons are essentially stationary. The profile of their distribution can be deduced from the energy profile of the polaritons, and largely resembles the profile of the pump spot. However, the details of the distribution matter, especially when attempting to measure the strength of the interactions when using a non-resonant pump. To do so, one must isolate a sufficiently large population of observable polaritons far from all undetectable excitons. This is necessary so that measured interactions are caused only by detected particles, allowing the density dependence to be determined. One recent work \cite{SunNatPhys2017}, which implemented non-resonant excitation, used the assumption that all of the excitons not detected by the NA of the optics  had the low diffusion lengths reported in semiconductor QWs \cite{Heller1996, Ramsteiner1997, Zhao2002} and organics \cite{Akselrod2010, Akselrod2014} of $\lesssim1~\mu$m. To the contrary, we show that the bottleneck excitons do not exhibit the nearly stationary behavior of bare excitons, and yet can have a significant population. This can be used, at least in part, to explain the large discrepancy in measured values of the polariton-polariton interaction strength \cite{Rodriguez2016, Walker2017, SunNatPhys2017}.

\section{Experimental Methods}
\label{sec:experimentalMethods}

\begin{figure}[htbp]
\centering
\includegraphics[width=\linewidth]{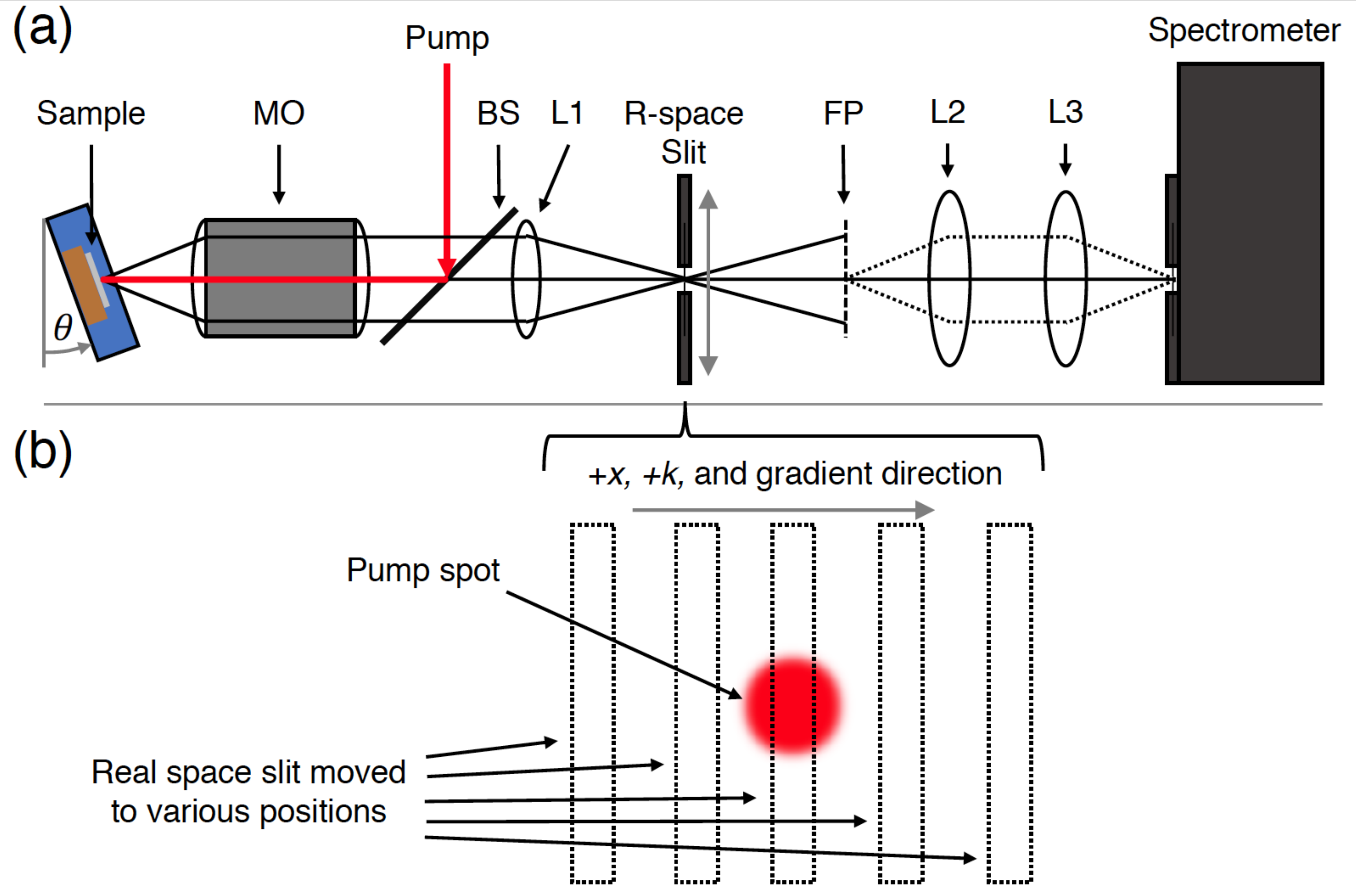}
\caption{(a) A diagram showing the basics of the optical setup, viewed from above. The sample was mounted in a cold-finger cryostat, which could be rotated by angle $\theta$. The pump laser was reflected off a beam splitter (BS) through the imaging microscope objective (MO). The PL was collected by the objective and then imaged by lens L1 onto a secondary real-space plane. At this plane, a movable slit (R-space Slit) was placed to select regions of the sample from which to resolve PL. The Fourier plane of the objective was also imaged by lens L1 at location FP. Lenses L2 and L3 then imaged this secondary Fourier plane onto the slit of the spectrometer. (b) A diagram of the real-space plane at the location of the slit (R-space Slit in (a)) as viewed along the imaging axis. The slit could be moved horizontally to select different regions of the image without changing the pump location. The $+x, +k$, and cavity gradient (``uphill'') directions are all the same.}
\label{fig:Diagram}
\end{figure}

The sample used in this experiment is the same as those used in previous work \cite{Nelsen2013, Steger2015, Liu2015, Myers2017, SunNatPhys2017, Ozden2018}. It consists of 7~nm GaAs QWs with AlAs barriers embedded within a distributed Bragg reflector (DBR) microcavity. The DBRs are made of alternating layers of AlAs and Al$_\mathrm{0.2}$Ga$_\mathrm{0.8}$As, with 32 periods in the top DBR and 40 in the bottom. The QWs are in sets of 4, with one set at each of the three antinodes of the 3$\lambda$/2 cavity. The sample exhibits polariton lifetimes near resonance of $\sim200$~ps. Because of a wedge in cavity length, the cavity energy changes across the sample. This gives a LP energy gradient pointing in the direction $8.6^\circ$ from the +x-direction as defined in Figure \ref{fig:Diagram}. The magnitude of the gradient was measured to be $\approx 5.2$~ meV/mm at the most excitonic detuning ($\delta \approx 7$~meV) used in this study. 

To produce polaritons, the sample was pumped non-resonantly (pump energy was 1726.8~meV) through the imaging objective lens, producing a pump spot with $\approx 3~\mu$m FWHM (see the Supplementary Information for details). The sample was held at $\sim 5$~K within a cryostat, which could be rotated around an axis perpendicular to the optical axis of the imaging objective. The numerical aperture of the objective was 0.40. This allowed collection of emission angles of about $-3^\circ$ to $43^\circ$ with the cryostat rotated to $20^\circ$, and up to $64^\circ$ with the cryostat rotated to $40^\circ$. A narrow slit was placed at a secondary real space imaging plane, which could be adjusted to select various regions of the image for angle-resolved imaging (see Figure \ref{fig:Diagram} for details of the setup).

With the pump tightly focused, the real space slit was swept through the region containing the pump spot while a CCD collected angle- and energy-resolved images of the photoluminescence (PL). This was done both near resonance ($\delta \approx 2$~meV) and at more excitonic detuning ($\delta \approx 8$~meV), using pump powers below the quasi-condensate threshold power, referred to as $P_\mathrm{th}$ throughout this work (see Ref. \onlinecite{MyersPRB2018} for a discussion of the various thresholds in these polariton systems).
For the angle-resolved images, the spectrometer slit selected a slice of the $k_\|$ plane along the x-axis. The slit was closed to $40~\mu$m for all of the images. This corresponded to a selection width of $0.034 \pm 0.004~\mu\mathrm{m}^{-1}$ in the $k_\|$ plane.

\section{Experimental Results}
\label{sec:data}

\begin{figure}[htbp]
\centering
\includegraphics[width=\linewidth]{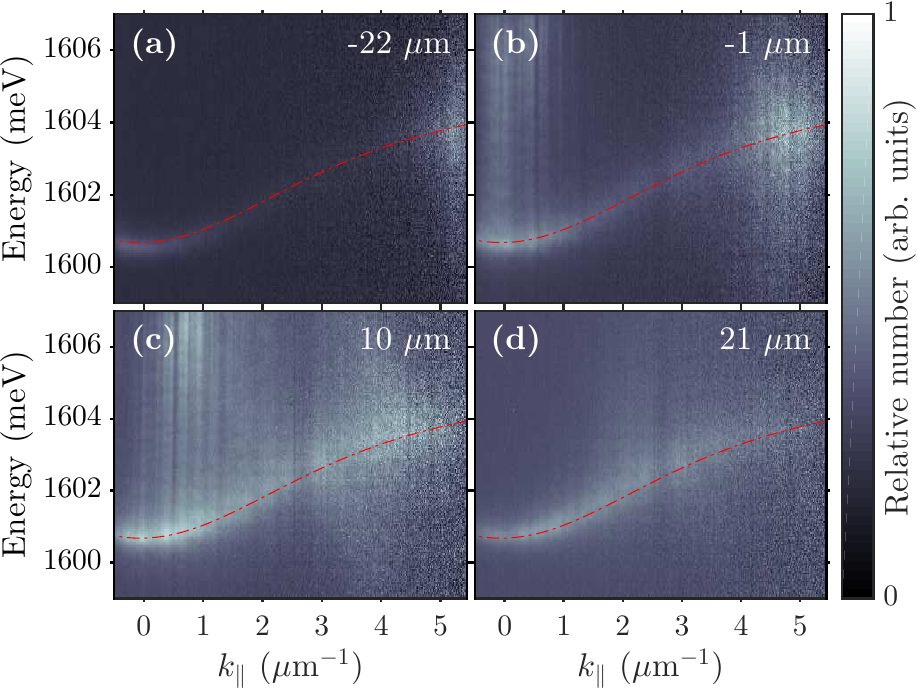}
\caption{Normalized lower polariton population as a function of energy and $k_\|$, taken from angle-resolved images and adjusted for the $k_\|$-dependent photon-fraction to show the relative particle populations. The pump power was about $P_\mathrm{th}/2$, and the detuning was about 8 meV. The positions of the real-space filter with respect to the pump spot are given in the upper right corners of each plot. The red lines show the theoretical LP dispersion. The counts for each image were normalized separately, so the counts of separate images are not comparable.}
\label{fig:EvK}
\end{figure}

Figure \ref{fig:EvK} shows examples of the energy- and angle-resolved data, collected with the sample rotated to an angle of $20^\circ$ to allow high angle imaging. Since the sample was held at an angle relative to the imaging objective, the range of collection angles is asymmetric. The data were adjusted at each $k_\|$ to give a relative number of particles from the PL intensity ($N_\mathrm{pol}  \equiv N_\mathrm{phot}\tau_\mathrm{pol}$). The PL intensity ($N_\mathrm{phot}$) is simply proportional to the intensity measured by the camera, while the LP lifetime depends mostly on the short cavity photon lifetime and the cavity fraction ($\tau_\mathrm{pol}\approx\frac{\tau_\mathrm{cav}}{f_\mathrm{cav}}$). Therefore, since the cavity lifetime is constant, the relative number of particles can be deduced from the PL intensity and cavity fraction at each $k_\|$.

\begin{figure}[htbp]
\centering
\includegraphics[width=\linewidth]{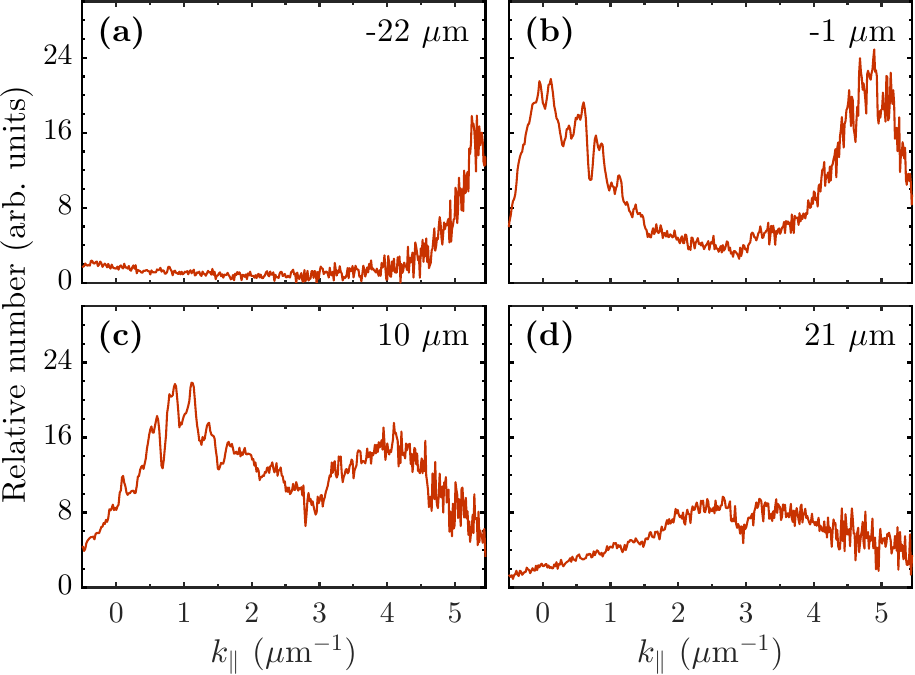}
\caption{Relative lower polariton population as a function of in-plane momentum ($k_\|$), derived from angle- and energy-resolved images integrated over energy. The inflection point of the LP dispersion is near $2.6~\mu\mathrm{m}^{-1}$. For these data, the sample was at $\theta = 20^\circ$ with respect to the imaging objective lens (as shown in Figure \ref{fig:Diagram}(a)), the pump power was about $P_\mathrm{th}/2$, and the detuning was about 8 meV. The positions of the real-space filter with respect to the pump spot are given in the upper right corners of each plot. The uncertainty of the relative number values is approximately the same as the scatter in the data.}
\label{fig:NvK}
\end{figure}

The LP dispersion in one direction is clearly visible, with a parabolic shape at low $k_\|$ and an inflection near $2.6~\mu\mathrm{m}^{-1}$. The dispersion flattens out at high $k_\|$ as the LP energy approaches the nearly flat exciton energy. Above the inflection, the energy line-width increases due to the high exciton fraction and the broader exciton energy line-width. Figure \ref{fig:NvK} shows the same data as Figure \ref{fig:EvK}, but integrated over energy, clearly revealing the $k_\|$ dependence of the populations within the LP band.

\section{Exciton Transport}
\label{sec:excitonTransport}

\begin{figure}[htbp]
\centering
\includegraphics[width=\linewidth]{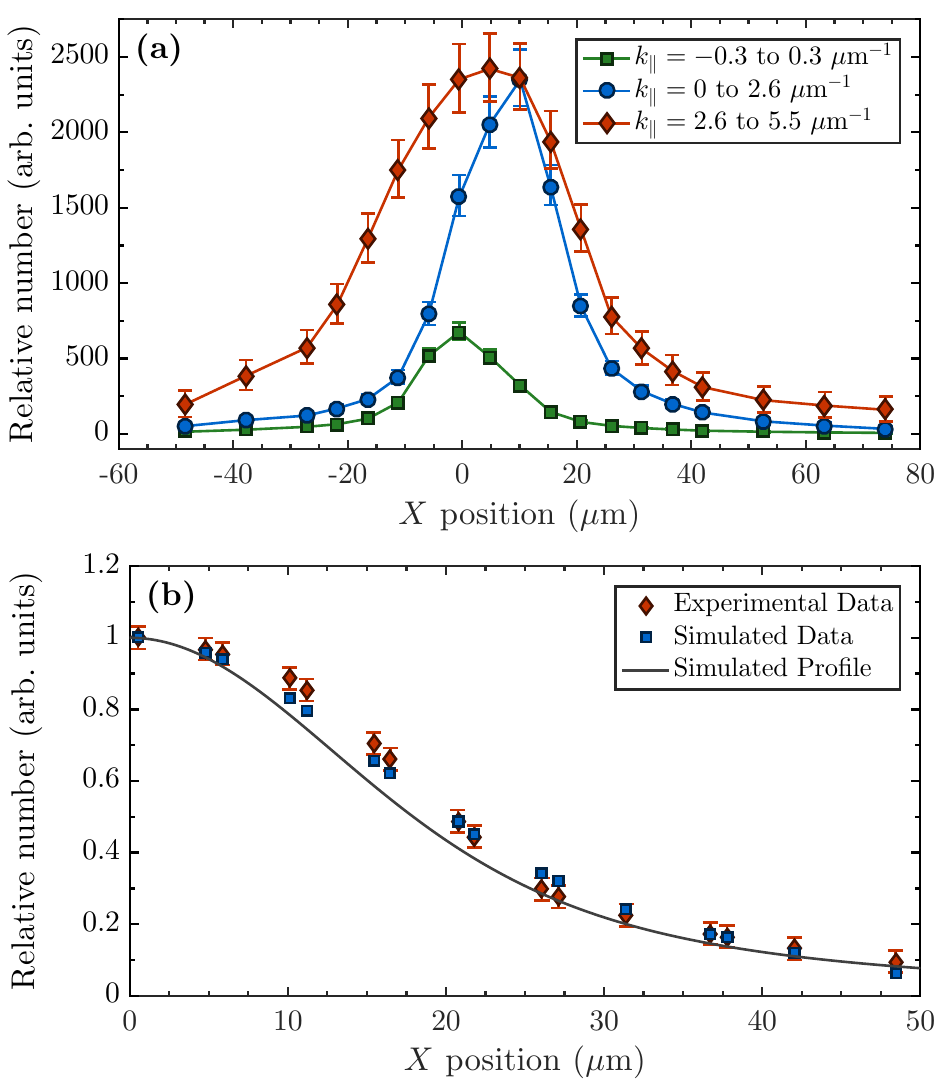}
\caption{(a) The relative number of particles within the LP band vs. position of the real space filter for various ranges of $k_\|$ for a pump power of about $P_\mathrm{th}/2$ and detuning of about 8 meV. The zero point in position was set by looking only at the symmetric range of visible $k_\|$ near $k_\|=0$ (green squares). (b) The experimental data (red diamonds) are the sum of the relative numbers at opposite sides of the pump spot (opposite $x$ positions) for the $k_\|$ range of 2.6 to 5.5~$\mu$m, shown as red diamonds in (a). The simulated profile (solid black line) is the Voigt profile representing the real-space distribution of the bottleneck excitons, which was normalized to show its shape compared to the data. The simulated data (blue squares) come from integrating the Voigt profile over small bounds in X, simulating the effect of the real-space slit in acquiring the experimental data. For details about the error bounds, see the Supplementary Information.}
\label{fig:NvX}
\end{figure}

Figure \ref{fig:NvX}(a) shows the relative populations for different ranges of in-plane momentum as the real-space filter is swept across the pump spot. This particular figure corresponds to polaritons and bottleneck excitons moving only in the +x-direction because of our choice of the range of collection angles. As expected for motion in this direction, the total number of ``normal'' polaritons below the inflection point in $k_\|$ peaks on the positive side of the pump spot, which was located by looking at the symmetric range of $k_\|$ near $k_\| = 0$. The relative number of bottleneck excitons with momentum above the inflection point peaks closer to the pump spot, but is broader overall. One feature to note is that there is asymmetry around the peak, with higher relative counts on the left side compared to the polaritons below the inflection, corresponding to back-scattered particles moving back toward the pump spot. This population of backward-moving, highly excitonic polaritons is visible in Figures \ref{fig:EvK}(a) and \ref{fig:NvK}(a) and is discussed below in Section \ref{sec:backscattering}. 

As mentioned above, we consider all of the emission from above the inflection to be part of the bottleneck exciton population. These bottleneck excitons, even at the resonant point for $k_\| = 0$, have excitonic fractions of at least 0.70, and are often undetected, e.g., in experiments like Ref.~\onlinecite{SunNatPhys2017}, due to their high emission angle ($\gtrsim20^\circ$). We note, however, that as seen in Figure~\ref{fig:NvX}(a), the population of these high-momenta bottleneck excitons is comparable to that of the entire thermalized population at low momentum. Although the occupation of these high-energy states is strongly suppressed in equilibrium by the Boltzmann factor $e^{-\Delta/k_BT}$, the well-known polariton bottleneck effect \cite{Askary1985} prevents full thermalization due to the suppressed phonon emission rate for excitons in these states. At higher densities, stimulated collisional effects can thermalize the polariton gas much more effectively \cite{Deng2010}; the experiments reported here were performed with excitation intensities well below the critical density threshold $P_{\rm thres}$ for Bose-Einstein condensation of polaritons, so that such collisional thermalization is suppressed, similar to the low-density conditions of Ref.~\onlinecite{SunNatPhys2017}.

We can assume that the contribution of the cavity gradient in the relatively short distances shown is negligible. Therefore, due to symmetry, the particles with positive $k_\|$ on the negative side of the pump spot can be assumed to have the same distribution in space as the particles with negative $k_\|$ on the positive side of the pump spot. This is utilized in Figure \ref{fig:NvX}(b), which simply adds the two sides of the positive $k_\|$ distribution together at equal distances. This estimates the full population integrated over both momentum directions. 

Given the circular symmetry of the experiment, we can also assume that the polaritons and bottleneck excitons move radially outward from the pump spot. This means that, for the slice of the $k_\|$ plane collected, only particles moving along a radial line parallel to that slice make a significant contribution to the measured population. Therefore, the experiment can be simulated with a 1D spatial distribution. We used a Voigt profile to produce this approximate distribution: 
\begin{equation}
N(x) = \int_{-\infty}^\infty \frac{A e^{-x'^2/2\sigma^2}}{(x-x'-x_0)^2+\gamma^{2}}dx'.
\label{eq:VoigtSI}
\end{equation}
By integrating this distribution with bounds similar to those given by the real-space slit width (typically about 20~$\mu$m), we were able to reconstruct the collected real-space integrated data. 

This method returned a FWHM for the distribution of the bottleneck excitons of $36\pm1~\mu$m at a pump power of $P_\mathrm{th}/2$. It is also mostly unaffected by pump power below threshold. This is much larger than the diffusion lengths for bare excitons in quantum wells, which, as discussed above, are typically of the order of 1~$\mu$m. This result also indicates that the reservoir exciton population far from the excitation region is not negligible, contrary to the assumption in Ref.~\onlinecite{SunNatPhys2017}. See the Supplementary Information for details and data for additional pump powers and detunings.

\section{Transport distance estimate}
\label{sec:transportEstimate}

\begin{figure}[htbp]
\centering
\includegraphics[width=\linewidth]{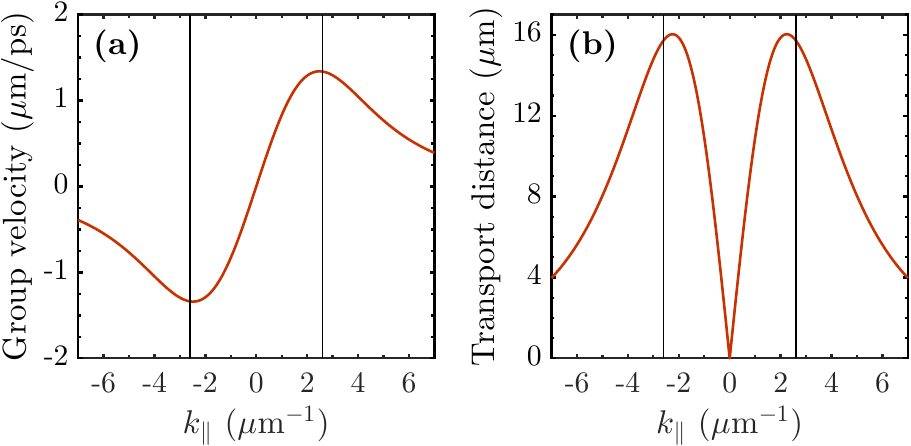}
\caption{(a) The group velocity of the lower polariton band for the actual sample parameters at a detuning of about 8 meV. (b) The estimated transport distance, assuming a cavity photon decay time of 100~ps and an exciton scattering time of 10~ps. The solid vertical lines mark $\pm2.6~\mu\mathrm{m}^{-1}$.}
\label{fig:transportEstimate}
\end{figure}

A simple calculation using the group velocity of the lower polariton band can be used to explain the overall effect. We assume that the exciton-polaritons we observe travel ballistically until they scatter out of the field of view of our detection or decay radiatively. The distance traveled is then approximately the group velocity of the particles times their effective time spent traveling ballistically. The group velocity can be easily calculated as $v_\mathrm{g} = \frac{1}{\hbar}\frac{dE}{dk_\|}$, which is plotted as a function of $k_\|$ in Figure \ref{fig:transportEstimate}(a), using the energy dispersion for $\delta \approx 8$~meV (shown in Figure \ref{fig:EvK}). A particle loss time ($\tau_\mathrm{loss}$) can be estimated by assuming loss rates from both radiative decay and scattering (since this experiment was done at low polariton density, the thermalization rate is assumed to be negligible). For the same reasons given in Section \ref{sec:excitonTransport}, we can treat scattering similarly to radiative loss, since any change of momentum away from the narrow slice in $k_\|$ that we collected for the experiment will result in the particle leaving the field of view. This leads to a loss time of $\tau_\mathrm{loss} = (1/\tau_\mathrm{r} + 1/\tau_\mathrm{s})^{-1}$, where $ \tau_\mathrm{r}$ is the radiative decay time and  $\tau_\mathrm{s}$ is the scattering time. 

As discussed in Section \ref{sec:data}, the radiative decay time is mostly dependent on the cavity photon decay time ($\tau_\mathrm{cav}$) and cavity fraction ($f_\mathrm{cav}$), giving $ \tau_\mathrm{r} \approx \frac{\tau_\mathrm{cav}}{f_\mathrm{cav}}$. The scattering time is similarly dependent on the exciton fraction ($f_\mathrm{exc}$), since the excitonic part is the primary part that undergoes scattering, giving $\tau_\mathrm{s} \approx \frac{\tau_\mathrm{s,exc}}{f_\mathrm{exc}}$, where $\tau_\mathrm{s,exc}$ is the exciton scattering time.

By multiplying the group velocity by the loss time, a transport distance can be estimated as 
\begin{equation}
\label{eq:transportDistance}
d \approx v_\mathrm{g} \tau_\mathrm{loss} \approx v_\mathrm{g} \bigg(\frac{f_\mathrm{cav}}{\tau_\mathrm{cav}} + \frac{f_\mathrm{exc}}{\tau_\mathrm{s,exc}} \bigg)^{-1},
\end{equation}
which is plotted in Figure \ref{fig:transportEstimate}(b), using a cavity photon decay time of 100~ps and an exciton scattering time of 10~ps. This plot shows that the distance is strongly peaked near the inflection points, with significant populations both above and below the inflection point traveling the longest distances. These parameters give an estimate consistent with our measured result of transport $\sim 20~\mu$m for both the bottleneck excitons and the polaritons.

\section{Backscattering at high in-plane momentum}
\label{sec:backscattering}

\begin{figure}[htbp]
\centering
\includegraphics[width=\linewidth]{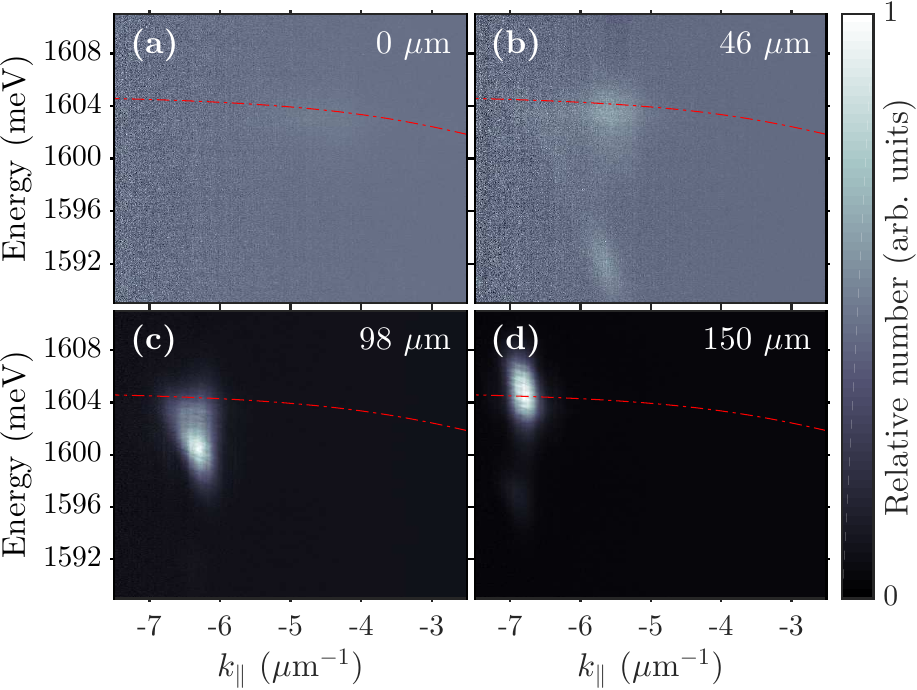}
\caption{The LP distribution as a function of energy and $k_\|$ at large angle (compare to Figure \ref{fig:EvK}). For these images, the sample was at $\theta = -40^\circ$ with respect to the imaging objective lens, the pump power was $\approx P_\mathrm{th}/2$, and the detuning was $\approx 7$~meV. The positions of the real-space filter with respect to the pump spot for each plot are given in the upper right corners. The red lines show the theoretical LP dispersion. The counts for each image were normalized separately, so the counts of separate images are not comparable.}
\label{fig:EvK_highK}
\end{figure}

As mentioned above, a significant population of bottleneck excitons can be seen with momentum in the direction back toward the pump spot. Figure \ref{fig:EvK_highK} shows data similar to those in Figure \ref{fig:EvK}, but with the sample rotated in the opposite direction and at a greater angle ($\theta = -40^\circ$). This allowed for collection up to $64^\circ$ external emission angle. Since these data are for negative momenta collected on the positive side of the pump spot, they correspond to backscattered bottleneck excitons traveling back toward the pump. They clearly show that this population is narrowly peaked and not a continuous distribution cut off by the numerical aperture of the optics. A similar population on the opposite side and with opposite momentum is also visible in Figure \ref{fig:EvK}(a).

The momentum of this backscattered population increases with increasing distance away from the pump spot. We do not fully understand this process, but we note that backscattered populations have been reported in similar samples under similar experimental conditions \cite{Ballarini2017}. This phenomenon could possibly be explained by the self interference of a population of polaritons with the positive and negative effective diffusive masses corresponding to momenta below and above the lower inflection point, respectively \cite{Colas2016}. It is also possible that bright soliton states could explain the observed backscattering. The bottleneck excitons populate the negative mass region of the LP dispersion that supports bright solitons \cite{EgorovSolitonsPRL2009}. Furthermore, backward flowing emission has been recently observed due to Cherenkov radiation from bright polarition solitons \cite{SkryabinNatComm2017}. While the present work differs in that we use non-resonant excitation, it is possible that bright solitons are still formed since the pumping populates states that support them.

\begin{figure}[htbp]
\centering
\includegraphics[width=\linewidth]{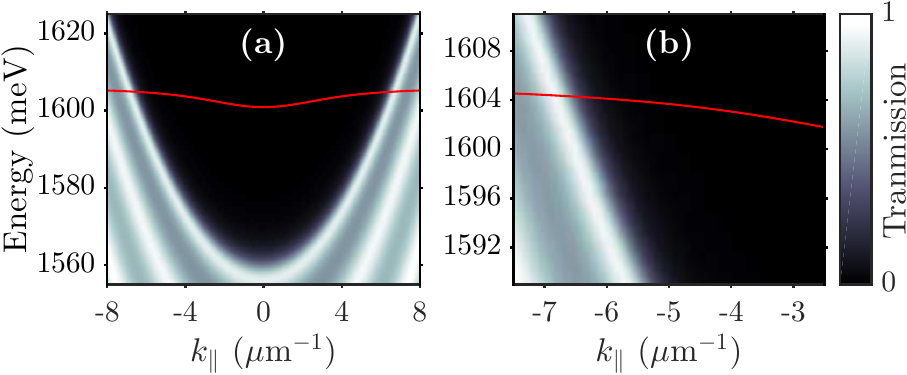}
\caption{The approximate transmission of the top DBR vs. energy and $k_\|$, produced using the transfer-matrix method. The red line shows the theoretical LP dispersion for a detuning of 7 meV, comparable to the red lines in Figure \ref{fig:EvK_highK}. (a) and (b) are identical except for the axes ranges.}
\label{fig:DBRdip}
\end{figure}

Another interesting effect can be seen in Figure \ref{fig:EvK_highK}: At large enough distance from the pump spot, the $k_\|$ of the population brings it near to where the distributed Bragg reflector (DBR) stop band dips cross the LP energy band (see Figure \ref{fig:DBRdip}). Figure \ref{fig:EvK_highK}(b) shows where the population begins to couple with the stop band dip, and Figure \ref{fig:EvK_highK}(c) shows the position at which it begins to heavily overlap. The PL intensity is greatly increased due to the new decay channel available to these highly excitonic polaritons, indicating a greatly decreased lifetime. Figure \ref{fig:EvK_highK}(d) shows the point where the peak momentum of the population matches the crossing point. The assumption that the LP lifetime is dependent upon the cavity fraction clearly breaks down for this $k_\|$ range, beginning around $5.5~\mu\mathrm{m}^{-1}$.

\section{Conclusions}

We have shown that excitonic polaritons usually considered part of the exciton reservoir exhibit much longer transport distances than previously expected. Rather than being almost entirely stationary, this portion of the exciton reservoir moves distances comparable to the much lighter and more photonic polaritons. We have also shown that there is a significant population of these excitonic polaritons with momentum backscattered toward the original pump spot. Since past work has clearly shown a mostly stationary exciton reservoir, this result indicates three separate categories are needed for particles in the LP mode: normal polaritons with relatively low $k_\|$, stationary reservoir excitons at very high $k_\|$, and highly mobile bottleneck excitons in between. These mobile excitons must be considered when attempting to isolate polaritons from the reservoir (see Appendix \ref{sec:ringPump} for an analysis of densities in a ring geometry). They also affect the potential profile felt by typically observed polaritons at low $k_\|$. As discussed above, this unexpectedly large exciton flow explains much of the blue shift of the polariton energy seen in Ref. \onlinecite{SunNatPhys2017}; the presence of many of these bottleneck excitons also can explain the large homogeneous line broadening seen in that study. In addition, recent studies \cite{PieczarkaArxiv2018} have shown that independent of any exciton flow, the increase of the barrier height in the traps used in Ref. \onlinecite{SunNatPhys2017} could lead to quantum confinement effects that can be dominant at low density and photonic detuning.

Long-range motion of excitons has been seen in other experiments, through different mechanisms. In the case of spatially indirect excitons in coupled quantum wells \cite{VorosPRL2005}, excitons moved over hundreds of microns due to their extraordinarily long lifetime, of the order of tens of microseconds. In other recent work \cite{LerarioLSA2017}, exciton-polaritons were seen to move hundreds of microns in a surface-wave geometry, in which the optical mode had no $k_\|=0$ state, but instead a built-in velocity of propagation. In the present case, the long-distance transport appears to be simply the result of longer-than-usual lifetime (due to the high cavity Q of our samples), low disorder, and the high velocity of the bottleneck excitons compared to the thermalized polaritons. Future work could be done to determine the details of the observed long-range transport and of the mobile excitons, as well as the origins of the backscattering effect. Potentially, the effect seen here of enhanced transport of excitons in a microcavity could be used to design room-temperature organic and inorganic devices such as solar cells with long-distance exciton transport.

\section*{Acknowledgements}

The work at Pittsburgh was funded by the Army Research Office (W911NF-15-1-0466). The work of sample fabrication at Princeton was funded by the Gordon and Betty Moore Foundation (GBMF-4420) and by the National Science Foundation MRSEC program through the Princeton Center for Complex Materials (DMR-0819860). S. M. also acknowledges the support of the Pittsburgh Quantum Institute.

\appendix

\section{Determining the Lower Polariton Detuning}
\label{sec:detuning}

In the long-lifetime sample used in this study, the upper polariton is not resolvable by directly imaging the PL, nor by reflectivity measurements. This makes determining the resonant position, coupling strength, and detuning more difficult than simply finding the point on the sample where the splitting between lower and upper polaritons is smallest. One method used to get around this difficulty is measuring the LP mass and the LP energy at various detunings. Because the cavity and other layer thicknesses are wedged by the growth process, varying the position on the sample also changes the detuning. Assuming that the exciton energy varies very little, the LP mass and energy can be used to find all of the other necessary parameters. However, it has large error bounds on the resonant position (resonant LP energy) and coupling strength, and therefore is not well suited to finding the absolute detuning values. This method was the primary one used in past work, and accounts for the parameters reported in those works, resulting in large uncertainty in the reported absolute detuning values (though the relative detunings are much more reliable).

A complementary method is photoluminescence excitation (PLE). Once the general vicinity of resonance is known, a PLE measurement can be used to find the upper polariton energy near resonance. This method is also imprecise for determining the resonant position, but provides a reliable and bounded measurement of the coupling strength. The full splitting between the lower and upper polariton bands at $k_\| = 0$ (typically called $\Omega$) was found to be 15.9 meV, with a lower bound of 15.2 meV and an upper bound of 16.7 meV.

For this study, the PLE measurement was combined with a method similar to the first. It differs in that, instead of finding the mass with a parabolic fit of the dispersion near $k_\| = 0$, a fit of the full theoretical LP dispersion was used at various detunings. By fitting the measured LP dispersion at a set of different detunings, accounting for the changes to the cavity and exciton energy due to the changing sample thicknesses, and using the coupling strength provided by PLE, all of the other necessary parameters can be determined. This method gives much more tightly constrained values for the detuning at any position on the sample than the more simple fit using the LP mass.

The overall change from previous work can be approximated by simply using 1597.3 meV as the resonant LP energy, rather than 1600.4 meV as previously reported. This shifts the detuning at the previously reported resonant position to $\delta \approx 8$~meV. This is the primary detuning used in this work. The other detuning in this work was previously reported as $\delta \approx -4$~meV, but is now reported as $\delta \approx 2$~meV. The previously reported energy splitting ($\Omega$) of 14.6 meV must also be replaced with the more precisely measured value of 15.9 meV.

\section{Application to Ring Pump Geometry}
\label{sec:ringPump}

The contribution of any LP above the inflection point was unaccounted for in Ref. \onlinecite{SunNatPhys2017} due to the use of an objective with NA = 0.28. In addition, the values for the cavity fraction, based on the detuning, were somewhat incorrect (see Appendix \ref{sec:detuning}). In this section, we will apply the measured exciton distribution above the inflection point to the ring pump geometry in order to help explain the extremely high interaction strength reported in that work. 

First, a fairly simple adjustment can be made to the reported exciton-exciton interaction ($g_\mathrm{xx}$) strength by considering the corrected detuning. The polariton-polariton interaction strength ($g_\mathrm{pp}$) can be related to the exciton-exciton interaction by the exciton fraction ($f_\mathrm{exc}$) of the polaritons of interest:
\begin{equation}
\label{eq:gppTOgxx}
g_\mathrm{pp} = (f_\mathrm{exc})^2 g_\mathrm{xx}.
\end{equation}
Additionally, the value for $g_\mathrm{pp}$ can be written as the ratio of the energy shift to the polariton density. The polariton density is found by relating the total number of emitted photons to the LP lifetime, which is in turn related to the cavity fraction ($f_\mathrm{cav}$) of the polaritons of interest (see Section \ref{sec:data}). Putting it all together gives
\begin{equation}
\label{eq:gettingGpp}
g_\mathrm{pp} = \frac{\Delta E}{n_\mathrm{pol}} = \frac{\Delta E}{n_\mathrm{ph}\frac{\tau_\mathrm{pol}}{\Delta t}} = \frac{\Delta E \Delta t f_\mathrm{cav}}{n_\mathrm{ph} \tau_\mathrm{cav}}
\end{equation}
where $\Delta E$ is the energy blue shift, $n_\mathrm{pol}$ is the polariton density, $n_\mathrm{ph}$ is the emitted photon density, $\Delta t$ is the integration time for photon detection, $\tau_\mathrm{pol}$ is the polariton lifetime, and $\tau_\mathrm{cav}$ is the cavity lifetime. Now if we consider a correction to the cavity and exciton fractions, and combine Equations \ref{eq:gppTOgxx} and \ref{eq:gettingGpp}, a corrected exciton-exciton interaction strength $g_\mathrm{xx}'$ can be expressed as 
\begin{equation}
\label{eq:correctedInteration}
g_\mathrm{xx}' = \frac{g_\mathrm{pp}'}{(f_\mathrm{exc}')^2} = \frac{g_\mathrm{pp}}{(f_\mathrm{exc}')^2 }\frac{f_\mathrm{cav}'}{f_\mathrm{cav}} = g_\mathrm{xx}\bigg(\frac{f_\mathrm{exc}}{f_\mathrm{exc}'}\bigg)^2\frac{f_\mathrm{cav}'}{f_\mathrm{cav}},
\end{equation}
where the primes indicate the corrected values. Applying this to the results in Ref. \onlinecite{SunNatPhys2017} brings the value for $g_\mathrm{xx}$ down from 1.74 meV$\mu\mathrm{m}^2$ to about 0.54 meV$\mu\mathrm{m}^2$. 

In order to correct for the excitons above the inflection, we must consider the two separate contributions to the energy blue shift by the polaritons and the bottleneck excitons:
\begin{equation}
\label{eq:blueShift}
\Delta E = \Delta E_\mathrm{pol} + \Delta E_\mathrm{exc} = g_\mathrm{pp}n_\mathrm{pol} + g_\mathrm{xx}f_\mathrm{exc}n_\mathrm{exc}.
\end{equation}
The effective exciton density of the bottleneck excitons is given by $n_\mathrm{exc}$, and has already been adjusted for their very high exciton fractions. In Ref. \onlinecite{SunNatPhys2017}, the exciton density was assumed to be negligible. Using this assumption, and the relation given in Equation \ref{eq:gppTOgxx}, we get
\begin{equation}
\label{eq:wrongGxx}
g_\mathrm{xx} = \frac{g_\mathrm{pp}}{(f_\mathrm{exc})^2} = \frac{\Delta E}{n_\mathrm{pol}}\frac{1}{(f_\mathrm{exc})^2},
\end{equation}
where the full energy shift is attributed to the measured polariton density. If instead we only attribute the shift from the polaritons to the polariton density, we get the corrected interaction strength
\begin{equation}
\label{eq:correctedGxxWork}
\begin{split}
g_\mathrm{xx}'& = \frac{\Delta E_\mathrm{pol}}{n_\mathrm{pol}(f_\mathrm{exc})^2}
\\& = \frac{\Delta E - \Delta E_\mathrm{exc}}{n_\mathrm{pol} (f_\mathrm{exc})^2}
\\& = \frac{\Delta E}{n_\mathrm{pol}(f_\mathrm{exc})^2} - \frac{g_\mathrm{xx}'f_\mathrm{exc}n_\mathrm{exc}}{n_\mathrm{pol}(f_\mathrm{exc})^2}
\\\Rightarrow g_\mathrm{xx}'\bigg(1 + \frac{n_\mathrm{exc}}{n_\mathrm{pol}f_\mathrm{exc}}\bigg)& = \frac{\Delta E}{n_\mathrm{pol}(f_\mathrm{exc})^2}.
\end{split}
\end{equation}
The last line in the above expression is simply the original $g_\mathrm{xx}$, which assumed a negligible exciton population. Thus, we can rearrange this for a simple relationship between the original and the corrected interaction strength values:
\begin{equation}
\label{eq:correctedGxxFinal}g_\mathrm{xx}' = g_\mathrm{xx}\bigg(1 + \frac{n_\mathrm{exc}}{n_\mathrm{pol}f_\mathrm{exc}}\bigg)^{-1}.
\end{equation}

\begin{figure}[tbp]
\centering
\includegraphics[width=\linewidth]{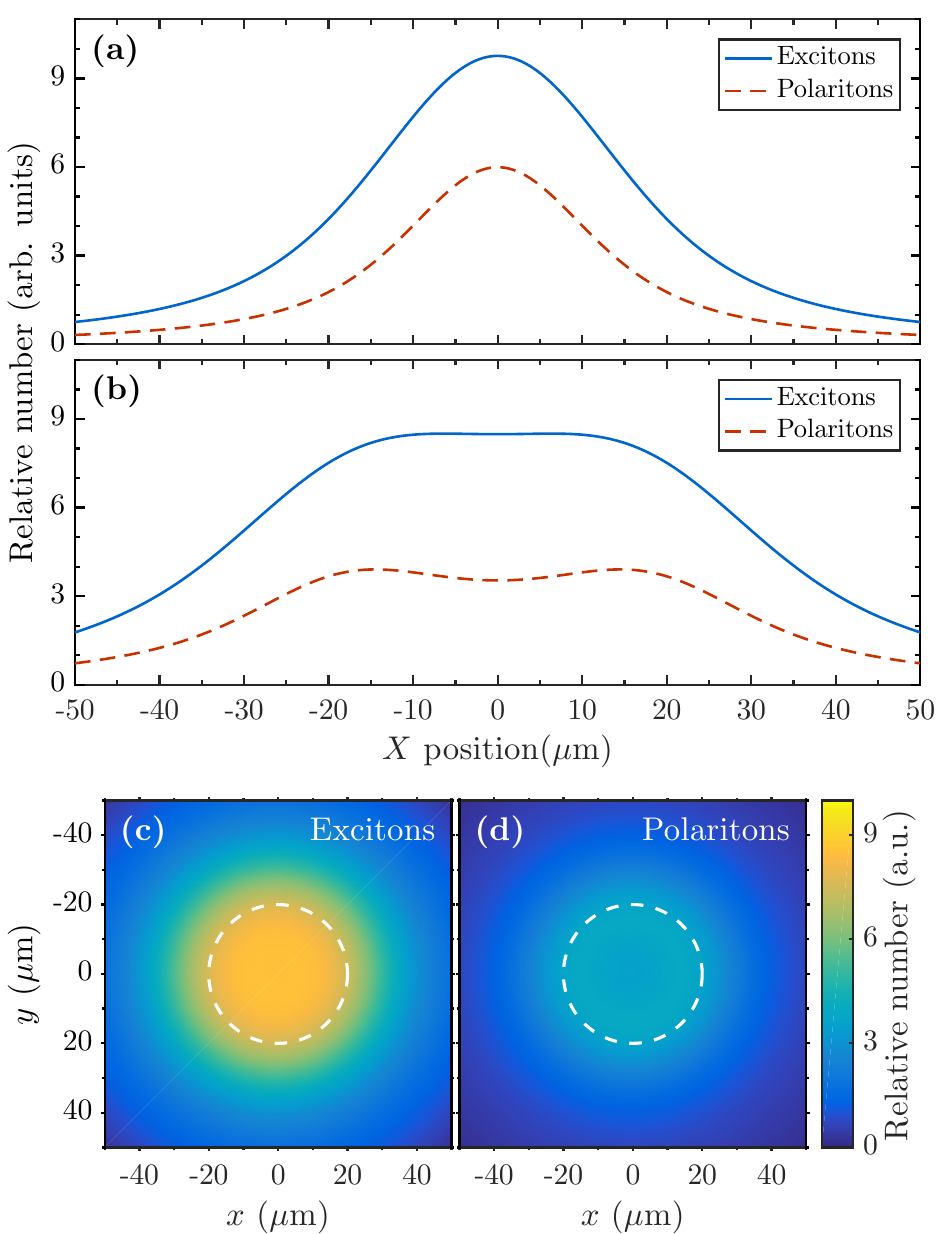}
\caption{(a) The profiles obtained by simulating the same experimental data used in Figure \ref{fig:NvX} for both the bottleneck excitons (above the inflection point) and the polaritons (below the inflection point). (b) A cross section of the resulting ring profile for both the excitons and the polaritons using a 40 $\mu$m diameter ring. (c) The exciton profile in two dimensions used to obtain the blue solid curve in (b). (d) The polariton profile in two dimensions used to obtain the red dashed curve in (b). The white dashed circles in both (c) and (d) mark the boundaries of the simulated ring. The relative numbers for all parts of this figure have been adjusted for the approximate exciton fractions, giving effective exciton numbers for each population.}
\label{fig:profiles}
\end{figure}

In order to find this correction, we only need to know the ratio of bottleneck excitons to polaritons at a given detuning in the ring geometry. To begin, we consider the profile for the exciton population shown in Figure \ref{fig:NvX}(b) and a similarly derived profile for the polaritons below the inflection. These profiles are both shown in Figure \ref{fig:profiles}(a), with the values adjusted for the approximate average exciton fractions of each population (taken to be 0.92 and 0.74 for the populations above and below the inflection, respectively).

\begin{figure}[htbp]
\centering
\includegraphics[width=\linewidth]{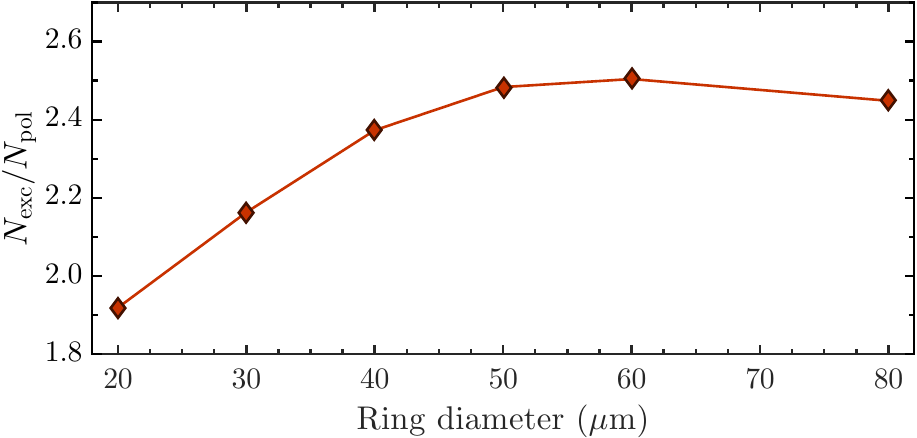}
\caption{The ratio of exciton (above the inflection point) to polariton (below the inflection point) number vs. ring diameter for distributions similar to and including those shown in Figure \ref{fig:profiles}(c-d). The numbers were both adjusted for the exciton fraction, giving a ratio of effective exciton populations. The numbers were obtained by integrating over a circle with 5 $\mu$m radius in the center of the ring profiles.}
\label{fig:ratioVdiameter}
\end{figure}

These profiles were found for what was essentially a point source, and give radial slices of the 2D profiles. The next step is to consider a series of radially symmetric profiles in two dimensions, each centered at even intervals around the perimeter of a circle, producing a ring. The resulting cross sections are given in Figure \ref{fig:profiles}(b) and the 2D profiles are given in Figure \ref{fig:profiles}(c) and (d), all for a ring with a diameter of 40 $\mu$m. By integrating a small region in the center for both the polariton and exciton reservoir distributions, we can extract a ratio of excitons to polaritons for various ring diameters (shown in Figure \ref{fig:ratioVdiameter}). By using the value for a 40 $\mu$m diameter ring, similar to that used in Ref. \onlinecite{SunNatPhys2017}, we get the ratio $\frac{n_\mathrm{exc}}{n_\mathrm{pol}f_\mathrm{exc}} = $ 2.4. By combining this correction with the detuning correction, we arrive at a new value for $g_\mathrm{xx}$ of 159~$\mu\mathrm{eV}\mu\mathrm{m}^2$. While this corrected value is still very high, it is at best an upper bound. In this work, we only carefully considered the exciton population up to $|k_\|| \simeq 5.5~\mu\mathrm{m}^{-1}$. Meanwhile we have also shown evidence of a backscattered population at much higher in-plane momentum. The existence of some excitons at these momenta at such large distances from the pump suggests that a population may be present at momenta greater than any that we probed in this work. In addition, other recent work \cite{PieczarkaArxiv2018} has shown that the quantum confinement from a ring trap leads to non-negligible blue shifts of the ground state in the center of the trap. Accounting for this effect would further decrease the reported interaction strength.

\section{Supplementary Information}

\subsection{Calculating Polariton Counts and the Error Bounds}
\label{sec:errorBounds}

The primary uncertainty for this work was in the measurement of the relative counts. This comes up in the processing of images similar to those shown in Figure 2 of the main text. This error comes from two main sources: the camera intensity value for each pixel and the cavity fraction. As mentioned above, the polariton counts were found by adjusting the camera intensity (counts) value for each pixel by the cavity fraction. In addition, the camera intensity values were also adjusted for variation in $\Delta k_\| = k_{\|,n} - k_{\|,n+1}$, where $k_{\|,n}$ corresponds to the $n^{th}$ pixel along the $k_{\|}$ axis of the image. This variation is caused inherently by the fact that $k_\|$ is not simply proportional to distance along the angle-resolved image. Instead, it is described by 
\begin{equation}
\label{eq:findingK}
k_\| = \frac{E}{\hbar c} \sin{\big(\arctan{\big(D/F\big)}\big)}
\end{equation}
where $E$ is the emitted photon energy, $D$ is the distance along the Fourier image from the $k_\| = 0$ point, and $F$ is a parameter related to the focal lengths and magnification of the optics ($F$ is most easily determined by calibration). Spherical aberration in the imaging optics also causes variation in $\Delta k_\|$ across the angle-resolved image. As a result, the relative polariton counts are given by
\begin{equation}
\label{eq:Counts}
N(k_\|) \sim \sum_{n}^{N_E} \frac{I_n(k_\|) }{\Delta k_\|(k_\|) f_\mathrm{cav}(k_\|)} = \frac{I(k_\|) }{\Delta k_\|(k_\|) f_\mathrm{cav}(k_\|)}
\end{equation}
where $N_E$ is the number of pixels summed up along the energy-axis, $I_n$ is the individual pixel intensity value, and $f_\mathrm{cav}$ is the cavity fraction of the polariton. Only the intensity varies significantly along the energy axis. Also, since we only care about relative counts, we only need the value to be proportional to the number of polaritons. The uncertainty in $\Delta k_\|$ is negligible due to small overall change and precise calibration. This leaves the uncertainty in the cavity fraction and the pixel intensity.

The pixel intensity error is related to random noise and the background subtraction. The random noise can be easily analyzed, giving a normal distribution with a standard deviation of 11 pixel counts (pixel counts being the numbers returned by the pixels). The background subtraction simply doubles this uncertainty, giving an overall error of 22 pixel counts. However, we have summed up many pixel intensity values at each $k_\|$ along the energy axis. This gives a new uncertainty of the overall intensity along the energy-axis of $\delta I = \delta I_n \sqrt{N_E}$. This value is constant in $k_\|$ for a fixed width along the energy-axis.

The uncertainty in cavity fraction is caused by uncertainty in the detuning at $k_\| = 0$, which is explained in Appendix A of the main text. The simplest way to determine the bounds on the cavity fraction is to repeat the calculation for the cavity fraction at each $k_\|$ at the maximum and minimum values given by the bounds of the detuning. The uncertainty in detuning is in turn mostly the result of the uncertainty in the coupling strength and the measurement of the LP energy at $k_\| = 0$. The upper and lower bounds of the coupling strength are propagated through the fitting process outlined in Appendix A of the main text, resulting in upper and lower bounds of the detuning as a function of $k_\|$ and the LP energy at $k_\| = 0$. The LP energy bounds are simply estimated as $\pm 0.2$~meV, based on the spectrally resolved images used in this study. The result is two new values for the upper and lower bounds of the cavity fraction, designated $f_\mathrm{cav,u}(k_\|)$ and $f_\mathrm{cav,l}(k_\|)$ respectively. Similarly, one can simply add and subtract the uncertainty to each intensity value, giving the upper and lower bounds $I_\mathrm{u}(k_\|)$ and $I_\mathrm{l}(k_\|)$.

Finally, we need to get the upper and lower bounds of the total polariton counts within a certain range of $k_\|$. One can simply use the combinations within the bounds that give the largest and smallest values. These are obviously
\begin{equation}
\label{eq:Bounds}
\begin{split}
&N_u \sim \sum_{n}\frac{I_\mathrm{u}(k_n)}{\Delta k_\|(k_n) f_\mathrm{cav, l}(k_n)}\\
\mathrm{and} &\\
&N_l \sim \sum_{n}\frac{I_\mathrm{l}(k_n)}{\Delta k_\|(k_n) f_\mathrm{cav, u}(k_n)}\\
\end{split}
\end{equation}
where $k_n$ is the $n^{th}$ $k_\|$ value, and the sum is over all the $k_\|$ values in the range of interest. These are the bounds shown in the figures with relative numbers of polaritons when comparing different ranges of $k_\|$ ({\it e.g.} the left column of Figure \ref{fig:NvX_additional}).

The process is slightly different when finding the error bounds for the data to determine the exciton transport length. Unlike the above described case, the relative counts between separate ranges of $k_\|$ are no longer important. Instead the only values that matter are relative counts at one position in real-space compared to another for the same $k_\|$ range. Since the cavity fraction as a function of $k_\|$ is the same for the entire set, the only important consideration is how the uncertainty of the cavity fraction affects the fractional change in counts from one position to the next. 

Following this reasoning, two contributions to the uncertainty were found for each position. The first was that due to the uncertainty in PL intensity, which was still relevant to each position individually and the same for all positions. The second was the fractional uncertainty from the possible variation in the cavity fraction. This latter uncertainty was found to be negligible compared to the uncertainty in PL intensity, but was still included in determining the error bounds. Since the two sides of the relative number vs. position data were added together, the uncertainty in PL intensity was doubled. Thus, the uncertainty was effectively twice the uncertainty due to PL intensity at each filter position. These are the bounds shown in figures with relative number of polaritons compared with the simulated data ({\it e.g.} the right column of Figure \ref{fig:NvX_additional}).

The error bounds for the FWHM values of the profiles of the excitonic bottleneck polaritons in real space were obtained by varying the parameters of the Voigt profile such that the simulated data remained mostly within the error bounds of the experimental data.

\subsection{Determining the Pump Spot Size}
\label{sec:pumpspot}

\begin{figure}[htbp]
\centering
\includegraphics[width=\linewidth]{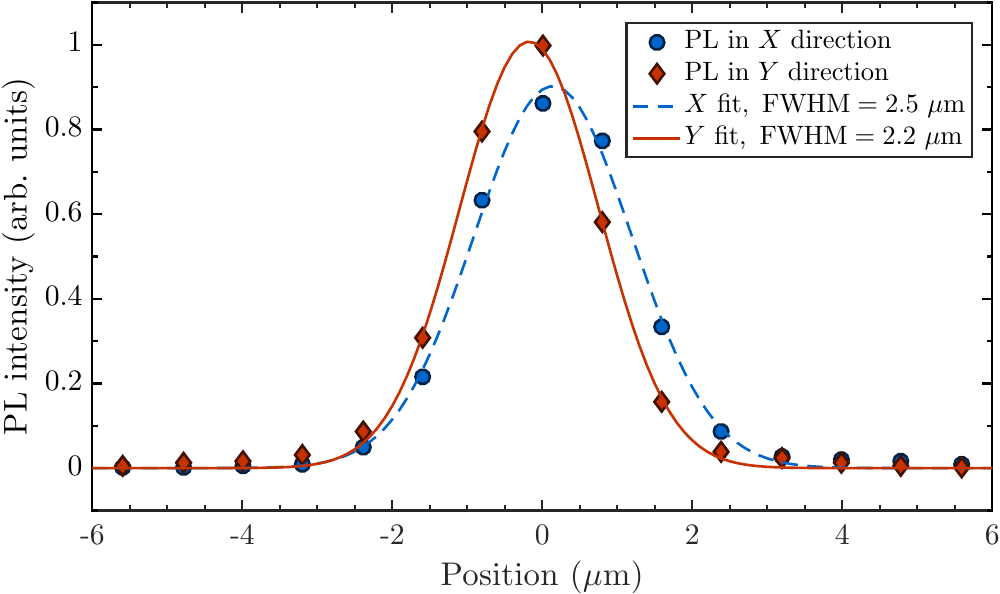}
\caption{PL intensity of the pump spot vs. position at the sample plane, in both X and Y directions. The fits are both Gaussian functions with FWHM $= 2\sqrt{2\log{2}}\sigma = 2.5~\mu$m and 2.2~$\mu$m in the $x$- and $y$-directions, respectively.}
\label{fig:pumpSpot}
\end{figure}

The pump spot size was determined by directly imaging the pump spot reflected from the sample plane (see Figure \ref{fig:pumpSpot}). The intensity profile was then fit by a Gaussian function. In both spatial directions, the fits gave a FWHM $\leq 2.5~\mu$m.

For this measurement, the pump spot was focused by looking directly at the sharp laser profile. However, during the experiment, the laser was focused by minimizing the real-space size of the PL. This was necessary because the laser was reflected from the sample at an angle greater than the numerical aperture of the objective during the actual experiment (see Figure 1(a) of the main text). While the pump spot size is minimized whenever the PL profile size is minimized, this is slightly less precise. Therefore, the above measurement of the pump spot size should be taken as both an approximate value and a lower bound. The pump spot size is thus given as $\approx 3~\mu$m. A measurement using hot luminescence at about 1701 meV gave essentially the same result.

\subsection{Additional Data Sets}
\label{sec:additonalData}

\begin{figure}[htbp]
\centering
\includegraphics[width=\linewidth]{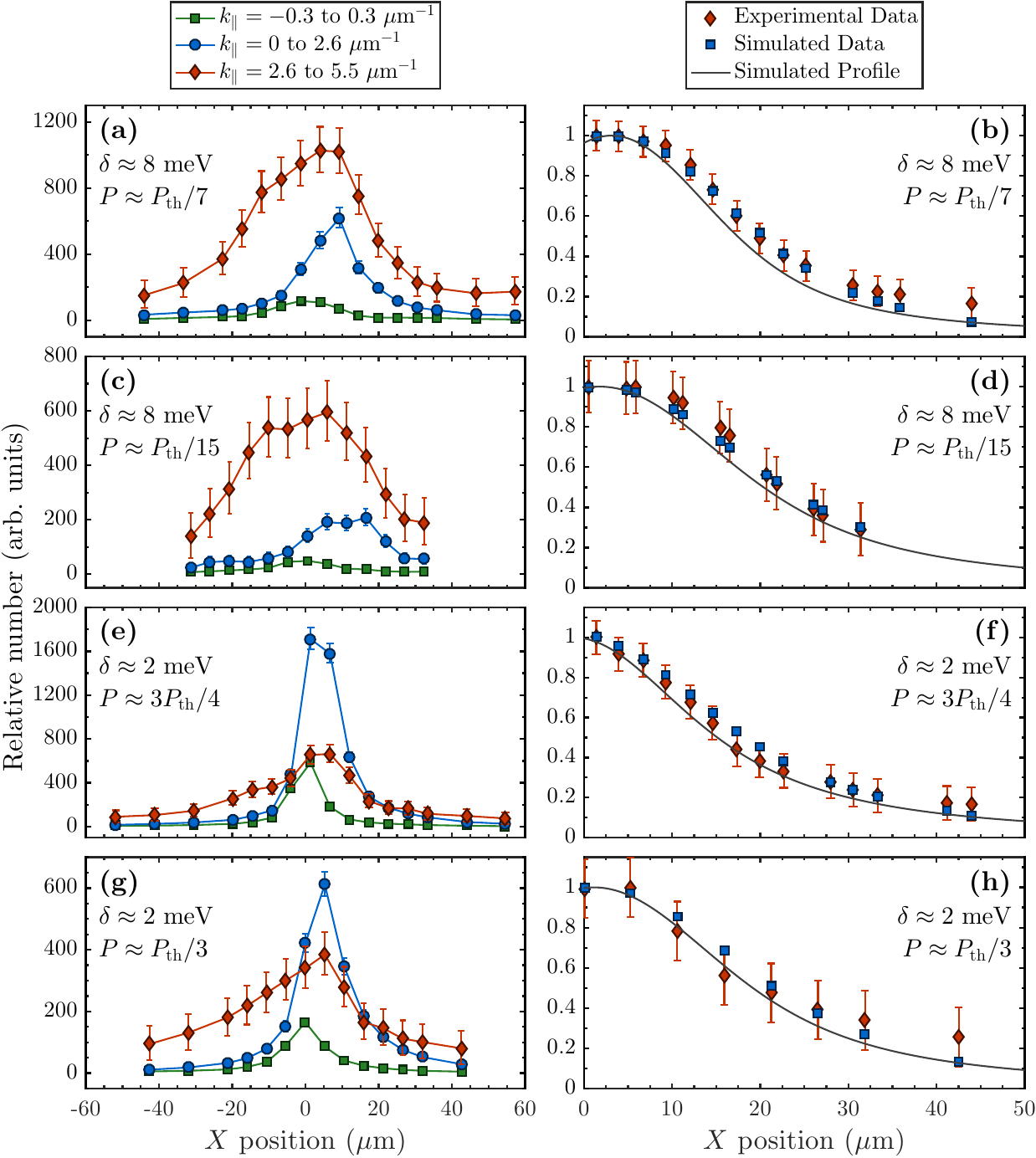}
\caption{Additional data similar to those shown in Figure 4 of the main text, but with different parameters, which are given in each plot. The experimental data for the plots in the right column in each case come from data with $k_\|$ from 2.6 to 5.5 $\mu$m.}
\label{fig:NvX_additional}
\end{figure}

Table \ref{tab:VoigtParameters} gives the various parameters for each data set and its corresponding profile. Figure \ref{fig:NvX_additional} shows plots of additional data sets similar to those in Figure 4 of the main text, along with the profiles used to produce the simulated data. 

\begin{table*}[htbp]
\centering
\begin{tabular}{C{0.85in}C{0.85in}C{0.85in}C{0.85in}C{0.85in}C{0.85in}C{0.9in}}
\hline
Figure & $\delta$ (meV) & Pump Power & $\sigma$ ($\mu$m) & $\gamma$ ($\mu$m) & $x_0$ ($\mu$m) & FWHM ($\mu$m)\\
\hline
4(b) & 8 & $P_\mathrm{th}/2$ & 8.7 & 11.8 & 0 & $36\pm1$\\
\ref{fig:NvX_additional}(b) & 8 & $P_\mathrm{th}/7$ & 7.1 & 9.2 & 3.1 & $29\pm4$\\
\ref{fig:NvX_additional}(d) & 8 & $P_\mathrm{th}/15$ & 7.7 & 13.9 & 1.8 & $37\pm9$\\
\ref{fig:NvX_additional}(f) & 2 & $3P_\mathrm{th}/4$ & 4.5 & 14.5 & -1.2 & $33\pm10$\\
\ref{fig:NvX_additional}(h) & 2 & $P_\mathrm{th}/3$ & 7.5 & 13.5 & 1.1 & $36\pm7$\\
\hline
\end{tabular}
\caption{The parameters for various data sets used in this study.}
\label{tab:VoigtParameters}
\end{table*}



%

\end{document}